\documentclass[11pt]{article}
\usepackage{geometry}

\usepackage{amsmath,amssymb,amsthm,booktabs}
\usepackage{mathrsfs}
\usepackage[OT1]{fontenc}
\usepackage[utf8]{inputenc}
\usepackage[colorlinks,citecolor=blue,urlcolor=blue]{hyperref}
\usepackage{txfonts}
\usepackage{indentfirst}
\usepackage{multirow}
\usepackage{float,subfig}

\usepackage{bm}
\usepackage{euscript}
\usepackage{graphicx}
\usepackage{multicol}
\usepackage[usenames,dvipsnames,svgnames,table]{xcolor}

\usepackage[round]{natbib}
\numberwithin{equation}{section} \theoremstyle{plain}
\newtheorem{theorem}{Theorem}[section]
\newtheorem{lemma}{Lemma}[section]

\newtheorem{proposition}{Proposition}[section]

\newtheorem{example}{Example}

\newtheorem{remark}{Remark}[section]

\numberwithin{equation}{section}

\begin{document}

\title{Bellman type strategy for the continuous time mean-variance model
\footnotemark[1]}
\author{Shuzhen Yang
\footnotemark[2]
\footnotemark[3]}

\renewcommand{\thefootnote}{\fnsymbol{footnote}}

\footnotetext[1]{

\textbf{\ \ Journal of Economic Literature classification Numbers}: G11, D81, C61.

\ \ \textbf{MSC2010 subject classification}: 91B28; 93E20; 49N10.
}
\footnotetext[2]{Shandong University-Zhong Tai Securities Institute for Financial Studies, Shandong University, PR China, (yangsz@sdu.edu.cn).}
\footnotetext[3]{This work was supported by the National Natural Science Foundation of China (Grant No.11701330) and Young Scholars Program of Shandong University.}

\date{}
\maketitle

\renewcommand{\thefootnote}{\alph{footnote}}

 \textbf{Abstract}: To investigate a time-consistent optimal strategy for the continuous time  mean-variance model, we develop a new method to establish the  Bellman principle. Based on this new method, we obtain a time-consistent dynamic optimal strategy that differs from the pre-committed and game-theoretic strategies. A comparison with the existing results on the continuous time mean-variance model shows that our method has several advantages. The explicit solutions of the dynamic optimal strategy and optimal wealth are given. When the dynamic optimal strategy is given at the initial time, we do not change it in the following investment time interval.

\addcontentsline{toc}{section}{\hspace*{1.8em}Abstract}

\newpage

\section{Introduction}
In the portfolio selection problem, we want to minimize the risk within a given expected return of the wealth. To solve this problem, \cite{Ma52,Ma59} proposed a mean-variance model in a single-period case. Then, \cite{Me72} solved this single-period problem analytically using mild assumptions. In the single-period mean-variance model, the mean and variance of wealth are used to represent its expected return and risk, respectively. Following this original single-period framework for the portfolio selection problem, many authors begin to consider related problems and the multi-period mean-variance model. Different from the single-period framework, the investor needs to optimize the multi-period objectives in the multi-period mean-variance model but not only optimizes the next period objective.

Furthermore, the discrete and continuous time mean-variance portfolio selection models have been proposed for the multi-period framework. \cite{R89} investigated a mean-variance model for one risky asset stock and  a bond with a constant risk-free rate in a continuous-time setting, in which the author focused on  minimizing the variance of the wealth at the terminal time under the constraint on mean value. \cite{BP98} considered the portfolio strategies that are mean-variance efficient when continuous rebalancing is allowed between the initial time and the terminal time.  \cite{LN00} employed the results of stochastic optimal control theory to solved a discrete-time multi-period mean-variance problem by embedding the original problem into a multi-objective optimization framework. Following the same idea in  \cite{LN00}, \cite{ZL00} investigated an optimal strategy and efficient frontier for the continuous-time mean-variance problem. In contrast, \cite{D88} proposed a cost-efficient approach to the optimal portfolio selection in a straightforward manner. Based on the cost-efficient approach, \cite{BS14} considered the problem of a mean-variance optimal portfolio in the presence of a benchmark. The optimal strategy and efficient frontier in the continuous-time mean-variance problem, which derived by the cost-efficient approach, is consistent with the results of \cite{ZL00}. Further extensions to the  mean-variance problem in continuous time include those with bankruptcy prohibition, transaction costs, and random parameters in complete and incomplete markets \citep{BJPZ05,DXZ10,LZ02,L04,X05}.

Based on a general mean-field framework,  \cite{AD11} considered the optimal control problem of a stochastic differential equation
of mean-field type, also be called McKean–Vlasov type equation. Employing  the related stochastic maximum principle to the mean-variance portfolio selection problem, \cite{AD11} obtained an optimal strategy which is coincided with that in \cite{ZL00}. \cite{L12} investigated an integral form stochastic maximum principle for general mean-field optimal control systems. As an application, a mean-field type linear quadratic stochastic control problem is solved. \cite{BDL11} established a general stochastic maximum principle for the stochastic differential equations of mean-field type. In addition, \cite{FL16} studied the continuous time mean-variance portfolio optimization problem and obtained the related pre-committed strategy using the mean field approach. \cite{PW17} considered the optimal control of general stochastic McKean-Vlasov equation and established the dynamic programming principle for the value function in the Wasserstein space of probability measures. In addition, the linear-quadratic stochastic McKean-Vlasov control problem and an interbank systemic risk model with common noise were investigated in  \cite{PW17}, further see \cite{PW18}. Recently, \cite{IP19} considered a robust continuous-time mean-variance portfolio selection problem where the model uncertainty
affects the covariance matrix of multiple risky assets. Furthermore, \cite{IP19} obtained the explicit solution for the optimal robust portfolio strategies in the case of uncertain volatilities, which is coincided with that in \cite{ZL00} and \cite{FL16}.

The optimal strategy in the aforementioned multi-period mean-variance framework is a pre-committed strategy that strengths the premise that the investor needs to follow the strategy given at the initial time. However, if the optimal strategy is not time-consistent, the investor may not obey this strategy in the following investment time interval. Here, the time-consistent means that the investor obtains the same strategy at any time during investment time interval. Thus, developing a dynamic time-consistent strategy for the mean-variance model in the continuous time framework is significant. From a game point of view, by directly defining a local maximum principle, a game-theoretic approach is investigated to address the mean-variance model in the multi-period case. Furthermore, by introducing an adjustment term in the objective,  \cite{BC10} adopted a dynamic method to study  the mean-variance model. \cite{HJZ12} formulated a general time-inconsistent stochastic linear-quadratic control problem, and defined an equilibrium instead of optimal control, further see \cite{Y12}. In addition, a pre-committed strategy for mean-variance model is given in \cite{HJZ12}. \cite{HCM07} considered the large population stochastic dynamic games and the Nash certainty equivalence based control laws. \cite{BSY16} studied the linear-quadratic mean field games via the adjoint equation approach, further see \cite{BSY13}. \cite{BMZ14} studied the mean-variance problem with state dependent risk aversion. \cite{BKM17} established a general framework to study the time-inconsistent stochastic control in the continuous time framework. In particular, \cite{DJSX19} proposed a dynamic mean-variance analysis for log returns within the game-theoretic approach.

Different from the aforementioned continuous time mean-variance framework, we develop a new method to study the multi-period mean-variance model via the dynamic programming principle in this study. Let $X^{\pi}_{t,x}(\cdot)$ denote the wealth of the investor in the investment time interval $[t,T]$ with the initial time $t$ and state $x$, where $\pi(\cdot)$ is the related strategy. The objective of the investor is to minimize the variance of the wealth $\mathrm{Var}[X^{\pi}_{t,x}(T)]$ within a given mean level constraint on $\mathbb{E}[X^{\pi}_{t,x}(T)]$. The question of this problem is that the objective $\mathrm{Var}[X^{\pi}_{t,x}(T)]$ does not satisfy the iterated-expectation property. Therefore, we cannot directly use the dynamic programming principle in the theory of stochastic optimal control to solve this continuous time mean-variance problem. Noting that $\mathrm{Var}[X^{\pi}_{t,x}(T)]=\mathbb{E}[\big(X^{\pi}_{t,x}(T)-\mathbb{E}[X^{\pi}_{t,x}(T)]\big)^2]$, the term  $\mathbb{E}[X^{\pi}_{t,x}(T)]$ in the formula of $\mathrm{Var}[X^{\pi}_{t,x}(T)]$ is the main gap when we investigate a Bellman principle for variance $\mathrm{Var}[X^{\pi}_{t,x}(T)]$. To bridge this gap, we use a deterministic process to represent the mean process $\mathbb{E}[X^{\pi}_{t,x}(\cdot)]$,  which is motivated by Example \ref{ex-1}. Therefore, we introduce a new deterministic process $Y^{\pi}_{t,y}(\cdot)$, which satisfies $Y^{\pi}_{t,y}(\cdot)=\mathbb{E}[X^{\pi}_{t,y}(\cdot)]$. The objective becomes $\mathbb{E}[(X^{\pi}_{t,x}(T)-Y^{\pi}_{t,y}(T))^2]$ within a given mean constraint on $\mathbb{E}[X^{\pi}_{t,x}(T)]$.

In this study, we want to consider the following objective cost functional:
\begin{equation}\label{in-cost}
\tilde{J}(t,x,y,\mu;\pi(\cdot))=\mu \mathbb{E}[\big(X^{\pi}_{t,x}(T)-Y^{\pi}_{t,y}(T)\big)^2]-Y^{\pi}_{t,y}(T).
\end{equation}
Note that the definition of the cost functional (\ref{in-cost}) allows us to separate the process $Y^{\pi}_{t,y}(\cdot)$ from the wealth's variance. Then, a value function $V^{\mu}(t,x,y)$ is defined by optimizing the objective cost functional (\ref{in-cost}). We can prove that the value function $V^{\mu}(t,x,y)$ satisfies a Bellman principle, and a related Hamilton-Jocabi-Bellman equation is derived. Through a series of analyses, we can obtain the explicit solution for the value function $V^{\mu}(t,x,y)$ and related optimal strategy with $x\neq y$. To solve the original problem, we extend the explicit solution of the value function $V^{\mu}(t,x,y)$ to the case $x=y$.
Furthermore, we find a time-consistent dynamic optimal strategy that differs from the existing strategies and compare our dynamic optimal strategy  with the pre-committed and game-theoretic strategies. {For notation simplicity, we use the game-theoretic strategy to denote the optimal strategy that is developed by the game-theoretic approach.}

The remainder of this paper is organized as follows. In Section 2, we formulate the continuous time mean-variance model. Then, in Section 3, we investigate an optimal strategy and  establish  a  dynamic time-consistent relationship between the mean and variance of the investor's wealth. In Section 4, following the main results of Section 3, we compare the mean, variance of the investor's wealth and the dynamic optimal strategy of our method with that of the pre-committed and game-theoretic strategies. In Section 5, we consider a general setting for the mean-variance model. Finally, we conclude  the  paper in Section 6.

\section{Motivation of a new Bellman principle }
In this section, we show the motivation of our Bellman principle for the classical continuous time mean-variance model using the following example.
\begin{example}\label{ex-1}
Let us consider a simple stochastic process:
 $$
 X_{t,x}(s)=x+b(s-t)+\sigma[W(s)-W(t)],\ t\leq s\leq T,
 $$
where $b,\sigma$ are constants, $T>0$, and $W(\cdot)$ is a standard Brownian motion. We consider the following value function:
$$
V_1(t,x)=\mathbb{E}[X_{t,x}(T)].
$$
Employing the Bellman principle to $V_1(\cdot)$, one obtains
$$
V_1(t,x)=\mathbb{E}[V_1(s,X_{t,x}(s))],\quad t\leq s\leq T.
$$
Thus, $V_1(t,x)$ satisfies the following partial differential equation (PDE):
\begin{eqnarray}\label{ex-eq-1}
\left\{\begin{array}
{ll}
&\!\!\!\!\! \displaystyle \partial_tV_1(t,x)+b\partial_xV_1(t,x)+\frac{1}{2}{\sigma}^2\partial^2_{xx}V_1(t,x)=0,\\
& \!\!\!\!\!V_1(T,x)=x,\quad 0\leq t<T.
\end{array}\right.
\end{eqnarray}
Based on PDE (\ref{ex-eq-1}), we can find an unique classical solution,
$$
V_1(t,x)=x+b(T-t),
$$
from which, we can see that the second-order term,
$$
\frac{1}{2}{\sigma}^2\partial^2_{xx}V_1(t,x)=0.
$$
Therefore, equation (\ref{ex-eq-1}) becomes
\begin{eqnarray}\label{ex-eq-2}
\left\{\begin{array}
{ll}
&\!\!\!\!\!\partial_tV_1(t,x)+b\partial_xV_1(t,x)=0,\\
& \!\!\!\!\!V_1(T,x)=x,\quad 0\leq t<T.
\end{array}\right.
\end{eqnarray}
These results motivate us to consider the expectation process of $X_{t,x}(\cdot)$,
$$
Y_{t,x}(s)=\mathbb{E}[X_{t,x}(s)]=x+b(s-t),\ t\leq s\leq T,
$$
and note that $V_1(t,x)=Y_{t,x}(T)=x+b(T-t)$ satisfies equation (\ref{ex-eq-2}).

In the following, we consider the value function of a nonlinear function of $\mathbb{E}[\cdot]$,
$$
V_2(t,x)=\Phi(\mathbb{E}[X_{t,x}(T)]),
$$
where $\Phi(x)$ has a continuous first-order derivative in $x\in \mathbb{R}$. Notice that we cannot use the Bellman principle for a nonlinear function of $\mathbb{E}[\cdot]$,  $\Phi(\mathbb{E}[X_{t,x}(T)])$. This is because the iterated-expectation property does not hold for $\Phi(\mathbb{E}[X_{t,x}(T)])$. Noting that, we can study the value function that is defined by process $Y_{t,x}(\cdot)=\mathbb{E}[X_{t,x}(\cdot)]$,
$$
V_2(t,x)=V_2(s,Y_{t,x}(s))=\Phi(Y_{t,x}(T)),\ t\leq s\leq T,
$$
and $V_2(t,x)$ satisfies the following equation:
\begin{eqnarray}\label{ex-eq-3}
\left\{\begin{array}
{ll}
&\!\!\!\!\!\partial_tV_2(t,x)+b\partial_xV_2(t,x)=0,\\
& \!\!\!\!\!V_2(T,x)=\Phi(x),\quad 0\leq t<T.
\end{array}\right.
\end{eqnarray}
\end{example}
\begin{remark}\label{re-1}  Example \ref{ex-1} indicates  that when we consider the value function of nonlinear function of $\mathbb{E}[X_{t,x}(\cdot)]$, we can introduce the process that denotes the expectation of state process $X_{t,x}(\cdot)$. Based on these observations, we  can establish the  Bellman principle for the value function through the mean process $\mathbb{E}[X_{t,x}(\cdot)]$. In the following, we use this idea to  study the mean-variance portfolio  problem in continuous time framework.
\end{remark}

\section{Bellman principle for mean-variance model }
\subsection{mean-variance model}
Given a complete
filtered probability space $(\Omega,\mathcal{F},P;\{ \mathcal{F}(s)\}_{s\geq
t})$, and $W(\cdot)$ is a $d$-dimensional standard Brownian motion defined on which with $W(t)=0$,  where $\{ \mathcal{F}(s)\}_{s\geq t}$ is the $P$-augmentation of the
natural filtration generated by  $W(\cdot)$. In the financial  market, we consider that one risk-free bond asset and $n$ risky  stock assets are traded, where the bond satisfies the following equation:
\begin{eqnarray*}
\left\{\begin{array}{rl}
\displaystyle \frac{\mathrm{d}S_0(s)}{S_0(s)} & \!\!\!= r(s)\mathrm{d}s,\;\;  \\
 S_0(t) & \!\!\!= s_0,\ \ t<s\leq T,
\end{array}\right.
\end{eqnarray*}
and the $i$'th ($1\leq i\leq n $) stock asset is described by
\begin{eqnarray*}
\left\{\begin{array}{rl} \displaystyle  \frac{\mathrm{d}S_i(s)}{S_i(s)} & \!\!\!=
b_i(s)\mathrm{d}t+ \displaystyle \sum_{j=1}^{{d}}\sigma_{ij}(s)\mathrm{d}W_{j}(s),\;\;
\\
 S_i(t) & \!\!\!= s_i,\ \ t<s\leq T,
\end{array}\right.
\end{eqnarray*}
where $r(\cdot)\in \mathbb{R}$ is the risk-free return rate of the bond, $b(\cdot)=(b_1(\cdot),\cdots,b_n(\cdot))\in{\mathbb{R}^n}$ is the expected return rate of the risky assets, and $\sigma(\cdot)=(\sigma_{1}(\cdot),\cdots,\sigma_{n}(\cdot))^{\top}\in \mathbb{R}^{n\times d} $ is the corresponding volatility matrix. Given initial capital $x>0$, $\displaystyle \gamma(\cdot)=(\gamma_1(\cdot),\cdots,\gamma_n(\cdot))\in \mathbb{R}^n$, where $\gamma_i(\cdot)=b_i(\cdot)-r(\cdot),\ 1 \leq i \leq n$. The investor's wealth $X^{\pi}_{t,x}(\cdot)$ satisfies
\begin{equation}\label{asset-1}
\left\{\begin{array}{rl}
\!\mathrm{d}X^{\pi}_{t,x}(s)  & \!\!\!=\big[r(s)X^{\pi}_{t,x}(s)  +\gamma(s)\pi(s)^{\top}   \big] \mathrm{d}s+\pi(s)\sigma(s)  \mathrm{d}W(s),  \\
\!X^{\pi}_{t,x}(t) & \!\!\!=x,\ \ t<s\leq T,
\end{array}\right.
\end{equation}
where $\pi(\cdot)=(\pi_1(\cdot),\cdots,\pi_n(\cdot))\in \mathbb{R}^{n}$ is the capital invested in the risky asset $S(\cdot)=(S_1(\cdot),\cdots,S_n(\cdot))\in \mathbb{R}^n$ and  $\pi_0(\cdot)$ is the capital invested in the bond. Thus, we have
$\displaystyle X^{\pi}_{t,x}(\cdot)=\sum_{i=0}^n\pi_i(\cdot)$.

In this study, we consider the following  mean-variance model:
\begin{equation}
J(t,x;\pi(\cdot))=\mathrm{Var}[X^{\pi}_{t,x}(T)]=
\mathbb{E}[\big{(}X^{\pi}_{t,x}(T)-\mathbb{E}[X^{\pi}_{t,x}(T)]\big{)}^2],\label{cost-1}%
\end{equation}
with the following constraint on the mean,
\begin{equation}
 \mathbb{E}[X^{\pi}_{t,x}(T)]=L. \label{mean-1}
\end{equation}
The set of admissible strategies $\pi(\cdot)$ is defined as:
$$
\mathcal{A}^T_t=\bigg{\{}\pi(\cdot):  \pi(\cdot)\in L^2_{\mathcal{F}}[t,T;\mathbb{R}^n]\bigg{\}},
$$
where $L^2_{\mathcal{F}}[t,T;\mathbb{R}^n]$ is the set of all square integrable measurable $\mathbb{R}^n$ valued  $\{\mathcal{F}_s\}_{s\geq t}$ adaptive processes. If there exists a  strategy $\pi^{*}(\cdot)\in \mathcal{A}^T_t$  that yields the minimum value of the cost functional (\ref{cost-1}), then we say that the  mean-variance model (\ref{cost-1}) is solved.

We suppose the following assumptions are used  to obtain the optimal strategy for the proposed  model (\ref{cost-1}):

{$\textbf{H}_1$}: $r(\cdot), b(\cdot)$ and $\sigma(\cdot)$ are bounded deterministic continuous functions.

{$\textbf{H}_2$}: $r(\cdot),\gamma(\cdot)>0$, $\sigma(\cdot)\sigma(\cdot)^{\top}>\delta  \textbf{I}$, where $\delta>0$ is a given constant and $\textbf{I}$ is the identity matrix of $\mathbb{S}^{n}$, and $\mathbb{S}^{n}$ is the set of symmetric matrices.

\subsection{Bellman principle}
In this section, we want to solve the mean-variance model via a  dynamic programming principle method. In detail, we set the term $\mathbb{E}[X^{\pi}_{t,x}(\cdot)]$ as a deterministic process $Y^{\pi}_{t,x}(\cdot)$, which differs from the stochastic term $X^{\pi}_{t,x}(\cdot)$. Therefore, we can establish the related Bellman principle. First, we introduce the following cost functional:
\begin{equation}
J(t,x,\mu;\pi(\cdot))=
\mu \mathrm{Var}[X^{\pi}_{t,x}(T)]-\mathbb{E}[X^{\pi}_{t,x}(T)],\label{cost-2}%
\end{equation}
where $\mu>0$ is the risk aversion coefficient and can be determined by the mean constraint $L$ in (\ref{mean-1}). Notice that,
$$
\mathrm{Var}[X^{\pi}_{t,x}(T)]=\mathbb{E}[\big{(}X^{\pi}_{t,x}(T)-\mathbb{E}[X^{\pi}_{t,x}(T)]\big{)}^2].
$$
However, we cannot obtain the Bellman principle for the term $[\mathbb{E}X^{\pi}_{t,x}(T)]^2$ because that $[\mathbb{E}(\cdot)]^2$ is a nonlinear function of $\mathbb{E}(\cdot)$. Remark \ref{re-1} suggests that we consider the dynamic programming principle for variables $(s,X^{\pi}_{t,x}(s),\mathbb{E}[X^{\pi}_{t,x}(s)]),\ t\leq s\leq T,\ x\in \mathbb{R}$. To separate the expectation term from the variance, we introduce the following auxiliary process $Y_{t,y}^{\pi}(\cdot)$, where $Y_{t,y}^{\pi}(\cdot)$ satisfies
 \begin{equation}\label{asset-2}
\left\{\begin{array}{rl}
\!\mathrm{d}Y^{\pi}_{t,y}(s)  & \!\!\!=\big[r(s)Y^{\pi}_{t,y}(s)  +\gamma(s)\mathbb{E}[\pi(s)^{\top}]   \big] \mathrm{d}s,  \\
\!Y^{\pi}_{t,y}(t) & \!\!\!=y,\ \ t<s\leq T.
\end{array}\right.
\end{equation}
Comparing  equations (\ref{asset-1}) and (\ref{asset-2}), we can see that
$$
Y^{\pi}_{t,y}(s)=\mathbb{E}[X^{\pi}_{t,y}(s)],\ \ t\leq s\leq T.
$$

Now, we introduce a useful version for cost functional (\ref{cost-2}):
\begin{equation*}
\begin{array}{rl}
&\tilde{J}(t,x,y,\mu;\pi(\cdot))\\
=&\mu \mathbb{E}[\big(X^{\pi}_{t,x}(T)-\mathbb{E}[X^{\pi}_{t,y}(T)]\big)^2]-\mathbb{E}[X^{\pi}_{t,y}(T)]\\
=& \mathbb{E}[\mu\big(X^{\pi}_{t,x}(T)-Y^{\pi}_{t,y}(T)\big)^2]-Y^{\pi}_{t,y}(T).
\end{array}
\end{equation*}
Obviously, we have
$$
\tilde{J}(t,x,x,\mu;\pi(\cdot))=J(t,x,\mu;\pi(\cdot)).
$$
Therefore, we consider the following value function:
\begin{equation}
V^{\mu}(t,x,y)=\inf_{\pi(\cdot)\in \mathcal{A}^T_t}\tilde{J}(t,x,y,\mu;\pi(\cdot)).\label{value-1}%
\end{equation}

\begin{remark}\label{re-2} In the definition of the cost functional $\tilde{J}(t,x,y,\mu;\pi(\cdot))$, we consider a stochastic process $X^{\pi}_{t,x}(\cdot)$ and  a deterministic process $Y^{\pi}_{t,y}(\cdot)$ under the same strategy $\pi(\cdot)\in \mathcal{A}^T_t$,  where $Y^{\pi}_{t,y}(\cdot)=\mathbb{E}[X^{\pi}_{t,y}(\cdot)]$. In this study, this relationship is useful. In the following, we derive the Bellman principle for the value function $V^{\mu}(t,x,y)$.
\end{remark}

Similar to the manner in Lemma 3.2, Chapter 4 in \cite{YZ99}, we obtain the following useful results.
\begin{lemma}\label{le-1}
Let Assumptions {$\textbf{H}_1$} and {$\textbf{H}_2$} hold. For any given $0\leq t\leq s\leq T,\ x,y\in \mathbb{R}$, $X^{\pi}_{t,x}(s)=\xi,\ X^{\pi}_{t,y}(s)=\eta\in L^2(\Omega)$, we have,
\begin{equation}\label{bel-1}
\tilde{J}(s,\xi,\mathbb{E}[\eta],\mu;\pi(\cdot))
= \mathbb{E}[\mu\big(X^{\pi}_{t,x}(T)-Y^{\pi}_{t,y}(T)\big)^2-Y^{\pi}_{t,y}(T)\ |\ \mathcal{F}_s].
\end{equation}
\end{lemma}
Based on Lemma \ref{le-1}, we have the following Bellman principle for the value function $V^{\mu}(t,x,y)$. The proofs of Theorems \ref{the-1} and \ref{the-2} are given in Appendix \ref{app}.
\begin{theorem}\label{the-1}
Let Assumptions {$\textbf{H}_1$} and {$\textbf{H}_2$} hold. For any given $0\leq t\leq s\leq T,\ x,y\in \mathbb{R}$, we have,
\begin{equation}\label{belm-1}
V^{\mu}(t,x,y)=\inf_{\pi(\cdot)\in \mathcal{A}_t^s}\mathbb{E}[V^{\mu}(s,X^{\pi}_{t,x}(s),Y^{\pi}_{t,y}(s))].
\end{equation}
\end{theorem}

\begin{theorem}\label{the-2}
Let Assumptions {$\textbf{H}_1$} and {$\textbf{H}_2$} hold. For any given $0\leq t\leq T,\ x,y\in \mathbb{R}$, $x\neq y$,
\begin{equation}\label{pde-10}
V^{\mu}(t,x,y)=\mu (x-y)^2e^{\int_t^T2r(h)\mathrm{d}h}-ye^{\int_t^Tr(h)\mathrm{d}h}
-\int_t^T\frac{\beta(h)}{4\mu}\mathrm{d}h,
\end{equation}
is the classical solution of the following partial differential equation (PDE),
 \begin{equation}\label{pde-0}
\left\{\begin{array}{rl}
  \!\!\! \partial_tV^{\mu}(t,x,y)=&\displaystyle-\inf_{\pi\in \mathbb{R}^n}
 \bigg{\{} \partial_xV^{\mu}(t,x,y)[r(t)x+\gamma(t)\pi^{\top}]
 +\partial_yV^{\mu}(t,x,y)[r(t)y+\gamma(t)\pi^{\top}]\\
 &+\displaystyle \frac{1}{2} \partial^2_{xx}V^{\mu}(t,x,y)\pi\sigma(t)\sigma(t)^{\top}\pi^{\top} \bigg{\}},  \\
  \!\!\! V^{\mu}(T,x,y)=& \mu (x-y)^2-y,
\end{array}\right.
\end{equation}
where $\beta(t)=\gamma(t)[\sigma(t)\sigma(t)^{\top}]^{-1}\gamma(t)^{\top}$, and the related optimal strategy is
$$
\pi^*(t,x,y)=\frac{1}{2\mu}\gamma(t)\big[\sigma(t)\sigma(t)^{\top}\big]^{-1}
e^{-\int_t^Tr(h)\mathrm{d}h},\ (t,x,y)\in [0,T]\times \mathbb{R}\times \mathbb{R}.
$$
\end{theorem}

\begin{remark}\label{re-00}
In Theorem \ref{the-2}, we obtain the explicit solution for $V^{\mu}(t,x,y)$ and the related optimal strategy $\pi^*(t,x,y)$ with $x\neq y$. The question is how to obtain the optimal strategy $\pi^*(t,x,y)$ with $x=y$. Note that, $V^{\mu}(t,x,y)$ and $\pi^*(t,x,y)$ are derived for $x\neq y$ and continuous on $(x,y)\in \mathbb{R}^2$. Thus, we extend the explicit solution of $V^{\mu}(t,x,y)$ and $\pi^*(t,x,y)$ to the case  $x\neq y$. In the following, we set
$$
V^{\mu}(t,x,x)=\lim_{y\to x}V^{\mu}(t,x,y),\quad \pi^*(t,x,x)=\lim_{y\to x}\pi^*(t,x,y).
$$

Thus, the original problem $J(t,x,\mu;\pi(\cdot))$ is solved based on the Bellman-type time-consistent optimal strategy $\pi^*(t,x,x),\ x\in \mathbb{R}$. In other words, we can obtain a
time-consistent optimal strategy for the original problem by extending the explicit solution of Hamilton system (\ref{pde-0}).
\end{remark}
\begin{remark}\label{re-0}
From Theorem \ref{the-2}, for given $(t,x,y)\in [0,T]\times \mathbb{R}\times \mathbb{R}$, we can obtain the optimal strategy $\pi^*(t,x,y)$, which deduces the optimal strategy at $(s,X^{\pi^*}_{t,x}(s),\mathbb{E}[Y^{\pi^*}_{t,y}(s)])$ is
$$
\pi^*(s,X^{\pi^*}_{t,x}(s),\mathbb{E}[Y^{\pi^*}_{t,y}(s)])=\frac{1}{2\mu}\gamma(s)\big[\sigma(s)\sigma(s)^{\top}\big]^{-1}
e^{-\int_s^Tr(h)\mathrm{d}h},
$$
which is independent from the initial state $(x,y)$. Thus, we omit the variable $(x,y)$ in $\pi^*(\cdot)$, and the optimal strategy
$$
\pi^*(s)=\frac{1}{2\mu}\gamma(s)\big[\sigma(s)\sigma(s)^{\top}\big]^{-1}
e^{-\int_s^Tr(h)\mathrm{d}h},\ t\leq s\leq T,
$$
does not change value at time $s>\max(t_1,t_2)$ with different initial times $t_1,t_2\geq 0$. Thus, we can see that $\pi^*(\cdot)$ is a time-consistent dynamic optimal strategy. Notice that, we have not shown how to determine the value of risk aversion parameter $\mu$. We need to use the mean level $L$ in constrained condition (\ref{mean-1}) to solve $\mu$, further see Remark \ref{re-9}.

\end{remark}
Based on Remark \ref{re-00}, let $y\to x$, combining the definition of $V^{\mu}(t,x,y)$, (\ref{value-1}) and explicit formulation of $V^{\mu}(t,x,y)$, (\ref{pde-10}), we set
\begin{equation}\label{pde-8}
\begin{array}{rl}
&V^{\mu}(t,x,x)\\
=&\lim_{y\to x}V^{\mu}(t,x,y)\\
=&\displaystyle\inf_{\pi(\cdot)\in \mathcal{A}^T_t}
\bigg{\{}\mu \mathrm{Var}[X^{\pi}_{t,x}(T)]-\mathbb{E}[X^{\pi}_{t,x}(T)]\bigg{\}}\\
=&\displaystyle -xe^{\int_t^Tr(h)\mathrm{d}h}
-\int_t^T\frac{\beta(h)}{4\mu}\mathrm{d}h.
\end{array}
\end{equation}

Note that in the first term $-xe^{\int_t^Tr(s)\mathrm{d}s}$ of the value function $V^{\mu}(t,x,x)$, the parameter $x>0$ is the initial wealth of the investor. The value function $V^{\mu}(t,x,x)$ is decreasing within $x\in (0,+\infty)$, which indicates that the large value of initial wealth brings small objective cost functional.
In general, we can assume a constant risk-free rate $r>0$, which shows that the first term of the value function $V^{\mu}(t,x,x)$ is decreasing with the length of the investment time interval $T-t$. In the second term $-\int_t^T\frac{\beta(h)}{4\mu}\mathrm{d}h$, $\beta(s)=\gamma(s)[\sigma(s)\sigma(s)^{\top}]^{-1}\gamma(s)^{\top},\ s\in [t,T]$. To clarify the effect of the second term, we consider a simple Black-Sholes setting, where $r,b,\sigma$ are independent from time $s\in[t,T]$ and $b_i>r>0, \sigma_{ij}=0,\ i\neq j,\ \sigma_{ii}=\sigma_i>0,\ 1\leq i,j\leq n$. Thus, we can obtain
$$
-\int_t^T\frac{\beta(h)}{4\mu}\mathrm{d}h=\frac{t-T}{4\mu}
\sum_{i=1}^n\bigg(\frac{b_i-r}{\sigma_i}\bigg)^2,
$$
where $\mu$ is the risk aversion parameter of the investor and $\displaystyle \frac{b_i-r}{\sigma_i}$ is the shape-ratio of the $i$'th risky asset. This formulation shows that cost functional $V^{\mu}(t,x,x)$ is decreasing with the risk aversion parameter $\mu$ and increasing with the shape-ratio of the risky asset. These results coincide with the  high return within high risk. Note that $\displaystyle \frac{b_i-r}{\sigma_i}>0,\ 1\leq i\leq n$; therefore, the cost functional $V^{\mu}(t,x,x)$ is increasing with the number of risky assets $n$, indicating that risk diversification may produce extra costs. In addition, the second term is decreasing with the length of the investment time interval $T-t$ which is same with the first term.

Note that the optimal strategy is given as follows:
$$
\pi^*(s)=\frac{1}{2\mu}\gamma(s)[\sigma(s)\sigma(s)^{\top}]^{-1}
e^{-\int_s^Tr(h)\mathrm{d}h}, \ t\leq s\leq T.
$$
Following this Black-Sholes setting, we have
\begin{equation}\label{op-1}
\pi^*(s)=\frac{e^{(s-T)r}}{2\mu}
(\frac{b_1-r}{\sigma_1^2},\frac{b_2-r}{\sigma_2^2},\cdots,\frac{b_n-r}{\sigma_n^2}), \ t\leq s\leq T.
\end{equation}
Thus, the investor invests an amount $\displaystyle \frac{e^{(s-T)r}}{2\mu}\frac{b_i-r}{\sigma_i^2}$ into the $i$'th risky asset and
an amount $\displaystyle x-\frac{e^{(s-T)r}}{2\mu}\sum_{i=1}^n\frac{b_i-r}{\sigma_i^2}$ into the risk-free asset. From the formulation of optimal strategy $\pi^*(\cdot)$, (\ref{op-1}), we can see that the optimal strategy $\pi^*(\cdot)$ is decreasing with the length of the investment time interval $T-s$ and decreasing with the risk aversion parameter $\mu$ which shows that the risk averse investor invests less money into the risky assets within a large value of the risk aversion parameter $\mu$. In addition, each element of $\pi^*(\cdot)$, $\displaystyle \frac{e^{(s-T)r}}{2\mu}\frac{b_i-r}{\sigma_i^2},\ 1\leq i\leq n$ is decreasing with the length of the investment time interval $T-s$ which indicates that the investor will add the proportion of the amount in the risky asset along with the holding time.
\begin{remark}\label{re-4}
We need to point out that the optimal strategy $\pi^*(\cdot)$ is independent from the wealth state. This finding coincides with the results in \cite{BC10}, in which the authors obtained an optimal strategy based on the game-theoretic approach. In fact, we may expect that an optimal strategy can depend on wealth $x$. However, we can solve this problem by changing the value of risk aversion parameter $\mu$. We can determine the value of $\mu$ using the initial time $t$ and wealth state $x$, and keep this risk aversion $\mu$ until the terminal time $T$. In addition, based on the formulation of the optimal cost functional $V^{\mu}(t,x,x)=\displaystyle -xe^{\int_t^Tr(h)\mathrm{d}h}
-\int_t^T\frac{\beta(h)}{4\mu}\mathrm{d}h$, we can take a large value of risk aversion $\mu$ and a large value of initial wealth $x$ to balance the cost functional $V^{\mu}(t,x,x)$. Further see \cite{BMZ14,BKM17} and \cite{DJSX19}.
\end{remark}

\subsection{Dynamic efficient frontier}
In this section, we want to derive the dynamic efficient frontier for $\mathbb{E}[X^{{\pi}^{*}}_{t,x}(s)]$ and $\mathrm{Var}[X^{{\pi}^{*}}_{t,x}(s)],\ t\leq s\leq T$. Plugging the optimal strategy
$$
\pi^*(s)=\frac{1}{2\mu}\gamma(s)\big[\sigma(s)\sigma(s)^{\top}\big]^{-1}
e^{-\int_s^Tr(h)\mathrm{d}h}, \ t\leq s\leq T,
$$
into the wealth equation (\ref{asset-1}), we can obtain that
$\mathbb{E}[{X}_{t,x}^{{\pi}^{*}}(\cdot)]$ and $\mathbb{E}[\big({X}_{t,x}^{{\pi}^{*}}(\cdot)\big)^2]$ satisfy the following linear ordinary differential equations.
\begin{equation}\label{step-asset-1}
\left\{\begin{array}{rl}
\!\!\!\mathrm{d}\mathbb{E}[X^{{\pi}^{*}}_{t,x}(s)]  & \!\!\!=\displaystyle\bigg[r(s)\mathbb{E}[X^{{\pi}^{*}}_{t,x}(s)]
+ \frac{1}{2\mu}e^{-\int_s^{T}r(h)\mathrm{d}h}\beta(s)  \bigg] \mathrm{d}s,  \\
 \!\!\!\mathbb{E}[X^{{\pi}^{*}}_{t,x}(t)] & \!\!\!=x,\ t< s\leq T,
\end{array}\right.
\end{equation}
and
\begin{equation}\label{step-asset-2}
\left\{\begin{array}{rl}
\!\!\!\mathrm{d}\mathbb{E}[\big(X^{{\pi}^{*}}_{t,x}(s)\big)^2]  & \!\!\!=\displaystyle\bigg[2r(s)\mathbb{E}[\big(X^{{\pi}^{*}}_{t,x}(s)\big)^2]+
  \frac{\mathbb{E}[X^{{\pi}^{*}}_{t,x}(s)] }{\mu}e^{-\int_s^{T}r(h)\mathrm{d}h}\beta(s)\\
  &+ \displaystyle \frac{1}{4\mu^2}e^{-\int_s^{T}2r(h)\mathrm{d}h}\beta(s)  \bigg] \mathrm{d}s,  \\
 \!\!\!\mathbb{E}[\big(X^{{\pi}^{*}}_{t,x}(t)\big)^2] & \!\!\!=x^2, \ t< s\leq T.
\end{array}\right.
\end{equation}
By equation (\ref{step-asset-1}), we have
\begin{equation}\label{step-asset-1-1}
\left\{\begin{array}{rl}
\!\!\!\mathrm{d}\big(\mathbb{E}[X^{{\pi}^{*}}_{t,x}(s)]\big)^2  & \!\!\!=\displaystyle\big[2r(s)\big(\mathbb{E}[X^{{\pi}^{*}}_{t,x}(s)]\big)^2
+ \frac{\mathbb{E}[X^{{\pi}^{*}}_{t,x}(s)]}{\mu}e^{-\int_s^{T}r(h)\mathrm{d}h}\beta(s)  \big] \mathrm{d}s,  \\
 \!\!\!\mathbb{E}[X^{{\pi}^{*}}_{t,x}(t)]^2 & \!\!\!=x^2,\ t< s\leq T,
\end{array}\right.
\end{equation}
Note that, $\mathrm{Var}[X^{{\pi}^{*}}_{t,x}(s)]=\mathbb{E}[\big(X^{{\pi}^{*}}_{t,x}(s)\big)^2] -\big(\mathbb{E}[X^{{\pi}^{*}}_{t,x}(s)]\big)^2,\ t\leq s\leq T$, combining equations (\ref{step-asset-2}) and (\ref{step-asset-1-1}), it follows that,
\begin{equation}\label{step-asset-3}
\left\{\begin{array}{rl}
\!\!\!\mathrm{d}\mathrm{Var}[X^{{\pi}^{*}}_{t,x}(s)]  & \!\!\!=\displaystyle\bigg[2r(s)\mathrm{Var}[X^{{\pi}^{*}}_{t,x}(s)] +
   \frac{1}{4\mu^2}e^{-\int_s^{T}2r(h)\mathrm{d}h}\beta(s)  \bigg] \mathrm{d}s,  \\
 \!\!\!\mathrm{Var}[X^{{\pi}^{*}}_{t,x}(t)] & \!\!\!=0, \ t< s\leq T.
\end{array}\right.
\end{equation}
From equations (\ref{step-asset-1}) and (\ref{step-asset-3}), for $t\leq s\leq T$, we can obtain $\mathbb{E}[X^{{\pi}^{*}}_{t,x}(s)]$ and $\mathrm{Var}[X^{{\pi}^{*}}_{t,x}(s)]$ as follows:
\begin{equation}\label{step-asset-4}
\left\{\begin{array}{rl}
\!\!\!\mathbb{E}[X^{{\pi}^{*}}_{t,x}(s)]&=\displaystyle  xe^{\int_t^sr(h)\mathrm{d}h}+e^{-\int_s^Tr(h)\mathrm{d}h}\int_t^s\frac{\beta(h)}{2\mu}\mathrm{d}h                       ,\\
\!\!\!\mathrm{Var}[X^{{\pi}^{*}}_{t,x}(s)]&=\displaystyle  e^{-\int_s^T2r(h)\mathrm{d}h}\int_t^s\frac{\beta(h)}{4\mu^2}\mathrm{d}h.
\end{array}\right.
\end{equation}

\begin{remark}\label{re-9}
Notice that, we introduce the risk aversion coefficient $\mu$ in cost functional (\ref{cost-2}). By equation (\ref{step-asset-4}), we can solve $\mu$ by constrained condition (\ref{mean-1}) as follows:
$$
\mu=\frac{\int_t^T\beta(h)\mathrm{d}h}{2\big(L- xe^{\int_t^Tr(h)\mathrm{d}h}\big)}.
$$
\end{remark}

From equation (\ref{step-asset-4}), for $t\leq s\leq T$, the relationship between $\mathbb{E}[X^{{\pi}^{*}}_{t,x}(s)]$ and $\mathrm{Var}[X^{{\pi}^{*}}_{t,x}(s)]$ is given as follows:
\begin{theorem}\label{the-2-1}
Let Assumptions {$\textbf{H}_1$} and {$\textbf{H}_2$} hold. We have
\begin{equation}\label{eff-1}
\displaystyle \mathrm{Var}[X^{{\pi}^{*}}_{t,x}(s)]=
\frac{\bigg(
\mathbb{E}[X^{{\pi}^{*}}_{t,x}(s)]-xe^{\int_t^sr(h)\mathrm{d}h}\bigg)^2}{\int_t^s\beta(h)\mathrm{d}h},\quad
t\leq s\leq T,
\end{equation}
where $\beta(h)=\gamma(h)[\sigma(h)\sigma(h)^{\top}]^{-1}\gamma(h)^{\top}, h\in [t,T]$.
\end{theorem}

\begin{remark}\label{re-5}
Based on equality (\ref{step-asset-4}), one obtains,
$$
\partial_s\mathbb{E}[X^{{\pi}^{*}}_{t,x}(s)]=xr(s)e^{\int_t^sr(h)\mathrm{d}h}
+r(s)e^{-\int_s^Tr(h)\mathrm{d}h}\int_t^s\frac{\beta(h)}{2\mu}\mathrm{d}h+\frac{\beta(s)}{2\mu}e^{-\int_s^Tr(h)\mathrm{d}h}>0
$$
and
$$
\partial_s\mathrm{Var}[X^{{\pi}^{*}}_{t,x}(s)]=2r(s)e^{-\int_s^T2r(h)\mathrm{d}h}\int_t^s\frac{\beta(h)}{4\mu^2}\mathrm{d}h
+\frac{\beta(s)}{4\mu^2}e^{-\int_s^T2r(h)\mathrm{d}h}>0.
$$
Thus, $\mathbb{E}[X^{{\pi}^{*}}_{t,x}(s)]$ and $\mathrm{Var}[X^{{\pi}^{*}}_{t,x}(s)]$ are increasing within $s\in [t,T]$.
Noting that $\mathbb{E}[X^{{\pi}^{*}}_{t,x}(s)]\geq xe^{\int_t^sr(h)\mathrm{d}h},\ s\in [t,T]$, $\mathrm{Var}[X^{{\pi}^{*}}_{t,x}(s)]$ is increasing with $\mathbb{E}[X^{{\pi}^{*}}_{t,x}(s)]$.
Furthermore, from formulation (\ref{eff-1}), we can see that the relationship between $\mathbb{E}[X^{{\pi}^{*}}_{t,x}(s)]$ and $\mathrm{Var}[X^{{\pi}^{*}}_{t,x}(s)]$ is uniformly for $s\in [t,T]$. This formulation is useful for the investor to check the relation between variance and mean value at each time $s\in [t,T]$.
\end{remark}

\section{Comparison with existence results}
In this section, we compare our dynamic optimal strategy with the existence results:  pre-committed and game-theoretic strategies. We focus on the properties of mean value, variance, optimal strategy and efficient frontier.
\subsection{Comparison with pre-committed strategy}
To solve the classical mean-variance model in the multi-period case,  \cite{LN00} considered the discrete-time multi-period mean-variance problem within a multi-objective optimization framework by  embedding the original problem into a stochastic linear-quadratic optimal control problem.  Based on the same idea in  \cite{LN00}, \cite{ZL00} formulated the continuous-time mean-variance problem  as a stochastic LQ optimal control problem. In contrast, \cite{D88} proposed a cost-efficient approach to solve the optimal portfolio selection in a straightforward manner. \cite{BS14} studied the problem of mean-variance optimal portfolio in the presence of a benchmark by the cost-efficient approach. Also, see \cite{AD11}, \cite{FL16} and \cite{IP19} for the pre-committed strategies.

Based on the same notation of this study, we review the main results of \cite{ZL00}. For the given initial time $t$ and state $x$, the optimal pre-committed strategy is given as follows:
\begin{equation}\label{pre-str-1}
{\pi}^{*}_1(s)=\gamma(s)[\sigma(s)\sigma(s)^{\top}]^{-1}[\lambda e^{-\int_s^{T}r(h)\mathrm{d}h}-X^{{\pi}_1^{*}}_{t,x}(s)],\quad  t\leq s\leq T,
\end{equation}
where $\displaystyle \lambda=\frac{e^{\int_t^T\beta(h)\mathrm{d}h}}{2\mu}+xe^{\int_t^Tr(h)\mathrm{d}h}$. The related efficient frontier is given as follows:
\begin{equation}\label{pre-eff}
\displaystyle  \mathrm{Var}[X^{{\pi}_1^{*}}_{t,x}(T)]=
\frac{\bigg{(}\mathbb{E}[X^{{\pi}_1^{*}}_{t,x}(T)]
-xe^{\int_t^{T}r(h)\mathrm{d}h}\bigg{)}^2}{e^{\int_t^{T}\beta(h)\mathrm{d}h}-1},
\end{equation}
where
$$
\displaystyle \mathbb{E}[{X}^{{\pi}^{*}_1}_{t,x}(s)]=xe^{\int_t^s[r(h)-\beta(h)]
\mathrm{d}h}+\lambda e^{-\int_s^{T}r(h)
\mathrm{d}h}[1-e^{-\int_t^{s}\beta(h)\mathrm{d}h}],\ t\leq s\leq T,
$$
and
$$
\displaystyle \mathbb{E}[{X}^{{\pi}^{*}_1}_{t,x}(T)]=
\frac{1}{2\mu}(e^{\int_t^{T}\beta(h)\mathrm{d}h}-1)+x e^{\int_t^{T}r(h)\mathrm{d}h}.
$$

In our model, by equality (\ref{step-asset-4}) in Subsection 3.2, we have
$$
\mathbb{E}[X^{{\pi}^{*}}_{t,x}(s)]=\displaystyle  xe^{\int_t^sr(h)\mathrm{d}h}+e^{-\int_s^Tr(h)\mathrm{d}h}\int_t^s\frac{\beta(h)}{2\mu}\mathrm{d}h,
$$
with the dynamic optimal strategy
$$
\pi^*(s)=\frac{1}{2\mu}\gamma(s)[\sigma(s)\sigma(s)^{\top}]^{-1}
e^{-\int_s^Tr(h)\mathrm{d}h}, \ t\leq s\leq T.
$$

By formula (\ref{pre-str-1}), the value of optimal pre-committed strategy $\pi^*_1(\cdot)$ at initial time $t$ is given as follows:
$$
\pi^*_1(t)=\frac{1}{2\mu}\gamma(t)[\sigma(t)\sigma(t)^{\top}]^{-1} e^{\int_t^{T}[\beta(h)-r(h)]\mathrm{d}h}.
$$
We have that $\pi^*(t)<\pi^*_1(t)$, where $\pi^*(t)<\pi^*_1(t)$ means that each element of $\pi^*(t)$ is smaller than that of $\pi^*_1(t)$. This is because the optimal pre-committed strategy only cares about the mean and variance at terminal time $T$, but not the entire investment time interval $[t,T]$. Thus, the optimal pre-committed strategy changes along with initial time $t$. Now, we return to our dynamic optimal strategy $\pi^*(\cdot)$ that can minimize the objective cost functional along the investment time interval  $[t,T]$. Thus, when we provide the dynamic optimal strategy $\pi^*(\cdot)$ at initial time $t$, then it will not change in the following time $s\in[t,T]$. Furthermore, we have the following properties of mean and variance under the optimal pre-committed strategy $\pi^*_1(\cdot)$ and the dynamic optimal strategy $\pi^*(\cdot)$. The proof of Proposition \ref{pro-1} is given in Appendix \ref{app}.
\begin{proposition}\label{pro-1}
For a given mean level $L>xe^{\int_t^{T}r(h)\mathrm{d}h}$ in the constrained condition (\ref{mean-1}), we have
\begin{equation}\label{pro-1-var}
\mathrm{Var}[X^{{\pi}^{*}}_{t,x}(T)]>\mathrm{Var}[X^{{\pi}_1^{*}}_{t,x}(T)].
\end{equation}
For a given risk aversion parameter $\mu>0$, one obtains
\begin{equation}\label{pro-1-var-1}
\mathrm{Var}[X^{{\pi}^{*}}_{t,x}(T)]<\mathrm{Var}[X^{{\pi}_1^{*}}_{t,x}(T)],\quad \mathbb{E}[{X}^{{\pi}^{*}}_{t,x}(T)]<\mathbb{E}[{X}^{{\pi}^{*}_1}_{t,x}(T)].
\end{equation}
\end{proposition}
\begin{remark}\label{re-6}
 For a given mean constrained value $L>xe^{\int_t^{T}r(h)\mathrm{d}h}$, $\mathbb{E}[{X}^{{\pi}^{*}}_{t,x}(T)]=\mathbb{E}[{X}^{{\pi}^{*}_1}_{t,x}(T)]=L$, based on the purpose of the dynamic optimal strategy $\pi^*(\cdot)$ and  the optimal pre-committed strategy $\pi^*_1(\cdot)$, we can see that the variance of the wealth ${X}^{{\pi}^{*}}_{t,x}(T)$ within dynamic optimal strategy $\pi^*(\cdot)$ is larger than the variance of the wealth ${X}^{{\pi}_1^{*}}_{t,x}(T)$ within optimal pre-committed strategy $\pi_1^*(\cdot)$. For a given risk aversion parameter $\mu>0$, the investor can obtain  smaller mean value and variance at terminal time $T$ within the dynamic optimal strategy $\pi^*(\cdot)$ than within the optimal pre-committed strategy $\pi^*_1(\cdot)$. Furthermore, for the given terminal time $T$, from the formulas of $\mathbb{E}[{X}^{{\pi}^{*}}_{t,x}(T)]$ and $\mathbb{E}[{X}^{{\pi}^{*}_1}_{t,x}(T)]$, we can see that the larger value of risk aversion parameter $\mu$  within a larger value of mean level $L$ in constrained condition (\ref{mean-1}).
\end{remark}

\subsection{Comparison with game-theoretic strategy}
Differ from the pre-committed strategies, by considering an adjustment term, \cite{BC10} adopted a dynamic method to study  the mean-variance model within a game-theoretic interpretation.  In contrast, based on the  game-theoretic approach, \cite{BMZ14,BKM17} studied the mean-variance problem with state dependent risk aversion.

Now, we introduce the results of Subsection 3.1 of \cite{BC10}. We assume that there is one bond with risk-free rate $r$ and one risky asset. The risky asset satisfies the constant elasticity of variance (CEV):
$$
\frac{\mathrm{d}S_1(s)}{S_1(s)}=b\mathrm{d}s+\sigma S_1^{\alpha}(s)\mathrm{d}W(s), \quad t\leq s\leq T,
$$
where $r,b,\sigma,\alpha$ are constants, $b>r>0,\sigma>0$. The optimal strategy $\pi^*_2(\cdot)$ in \cite{BC10} is given as follows:
\begin{equation}\label{cev-op}
\pi^*_2(s)=\frac{b-r}{2\mu\sigma^2S_1^{2\alpha}(s)}-\frac{1}{2\mu} \bigg(\frac{b-r}{\sigma S_1^{\alpha}(s)}\bigg)^2\frac{e^{-2\alpha r(T-s)}-1}{r}e^{-r(T-s)}, \quad t\leq s\leq T.
\end{equation}
Similar with the manner of Theorem \ref{the-2} and apply the results of Theorem \ref{the-2} to the CEV model, further see Theorem \ref{the-3}. The dynamic optimal strategy is given as follows:
\begin{equation}\label{cev-dy}
\pi^*(s)=\frac{b-r}{2\mu\sigma^2S_1^{2\alpha}(s)},\quad  t\leq s\leq T.
\end{equation}
\begin{remark}\label{re-7}
Note that if $\alpha=0$, the second term of optimal strategy $\pi^*_2(\cdot)$ is equal to $0$, thus, $\pi^*_2(s)=\pi^*(s),\ t\leq s\leq T$. This result demonstrates that our methodology developed in this study is a useful tool to establish a dynamic optimal strategy for the classical mean-variance model.

However, our method is different from that of \cite{BC10}. Note that, for $t\leq s\leq T$, when $S_1(s)>1$, we can obtain  $\pi^*_2(s)>\pi^*(s)$ for $\alpha>0$, and $\pi^*_2(s)<\pi^*(s)$ for $\alpha<0$. Compared with our dynamic optimal strategy $\pi^*(\cdot)$, the optimal strategy $\pi^*_2(\cdot)$  suggests that the investor  adds the investment  amount to the risky asset when the volatility of the risky asset becomes large, and reduces the investment amount to the risky asset when the volatility of the risky asset becomes small. In contrast, our dynamic optimal strategy $\pi^*(\cdot)$ suggests that the investor adds the investment amount to the risky asset when the volatility of the risky asset becomes small, and reduces the investment amount to the risky asset when the volatility of the risky asset becomes large. These results indicate that our dynamic optimal strategy $\pi^*(\cdot)$ is better than the optimal strategy $\pi^*_2(\cdot)$ that is derived based on the game-theoretic approach.
\end{remark}

\section{A general setting}
In this section, we consider the following general setting for the bond and the risky assets.
In the financial  market, there is one risk-free  bond asset and $n$ risky  stock assets that are traded, and the bond satisfies the following equation:
\begin{eqnarray*}
\left\{\begin{array}{rl}
\displaystyle \frac{\mathrm{d}P_0(s)}{P_0(s)} & \!\!\!= r(s,P_0(s))\mathrm{d}s,\;\;  \\
 P_0(t) & \!\!\!= p_0,\ \ t<s\leq T,
\end{array}\right.
\end{eqnarray*}
and the $i$'th ($1\leq i\leq n $) stock asset is described by
\begin{eqnarray*}
\left\{\begin{array}{rl} \displaystyle \frac{\mathrm{d}P_i(s)}{P_i(s)} & \!\!\!=
b_i(s,P_i(s))\mathrm{d}s+ \displaystyle  \sum_{j=1}^{{d}}\sigma_{ij}(s,P_i(s))\mathrm{d}W_{j}(s),\;\;
\\
 P_i(t) & \!\!\!= p_i,\ \ t<s\leq T,
\end{array}\right.
\end{eqnarray*}
 where $\sigma(\cdot)=(\sigma_{1}(\cdot),\cdots,\sigma_{n}(\cdot))^{\top}\in \mathbb{R}^{n\times d} $ is the corresponding volatility matrix. Given initial capital $x>0$, $\displaystyle \gamma(\cdot)=(\gamma_1(\cdot),\cdots,\gamma_n(\cdot))\in \mathbb{R}^n$, where $\gamma_i(\cdot)=b_i(\cdot)-r(\cdot),\ 1 \leq i \leq n$. The investor's wealth $X^{\pi}_{t,x}(\cdot)$ satisfies
\begin{equation}\label{g-asset-1}
\left\{\begin{array}{rl}
\!\mathrm{d}X^{\pi}_{t,x}(s)  & \!\!\!=\big[r(s,P_0(s))X^{\pi}_{t,x}(s)  +\gamma(s,P_0(s),P(s))\pi(s)^{\top}   \big] \mathrm{d}s\\
&+\pi(s)\sigma(s,P(s))  \mathrm{d}W(s),  \\
\!X^{\pi}_{t,x}(t) & \!\!\!=x,\ \ t<s\leq T,
\end{array}\right.
\end{equation}
where $\pi(\cdot)=(\pi_1(\cdot),\cdots,\pi_n(\cdot))\in \mathbb{R}^{n}$ is the capital invested in the risky assets, $P(\cdot)=(P_1(\cdot),\cdots,P_n(\cdot))\in \mathbb{R}^n$ and  $\pi_0(\cdot)$ is the capital invested in the bond.

We assume the following new Assumptions $\textbf{H}_3$ and $\textbf{H}_4$ for the above general setting.

{$\textbf{H}_3$}: For $(t,z)\in [0,T]\times \mathbb{R}$, $r(t,z) z,\ b(t,z)z$ and $\sigma(t,z)z$ are deterministic continuous functions and satisfy Lipschitz conditions in $z$.

{$\textbf{H}_4$}: $r(\cdot),\gamma(\cdot)>0$, $\sigma(\cdot)\sigma(\cdot)^{\top}>\delta  \textbf{I}$, where $\delta>0$ is a given constant and $\textbf{I}$ is the identity matrix of $\mathbb{S}^{n}$, and $\mathbb{S}^{n}$ is the set of symmetric matrices.

\bigskip

Notice that, Assumption $\textbf{H}_3$ is used to guarantee the existence and uniqueness of $P_0(\cdot)$ and $P(\cdot)$. Meanwhile, we will employ Assumption $\textbf{H}_4$ to obtain the optimal strategy. The main result of this section is given as follows and the proof is given in Appendix \ref{app}.
\begin{theorem}\label{the-3}
Let Assumptions {$\textbf{H}_3$} and {$\textbf{H}_4$} hold. For any given $0\leq t\leq T,\ x,y\in \mathbb{R}$, $x\neq y$,
\begin{equation}\label{g-pde-10}
V^{\mu}(t,x,y)=\mu (x-y)^2e^{\int_t^T2r(h,P_0(h))\mathrm{d}h}-ye^{\int_t^Tr(h,P_0(h))\mathrm{d}h}
-\int_t^T\frac{\mathbb{E}[\beta(h)]}{4\mu}\mathrm{d}h,
\end{equation}
is the classical solution of the following partial differential equation,
 \begin{equation}\label{g-pde-0}
\left\{\begin{array}{rl}
  \!\!\! \partial_tV^{\mu}(t,x,y)=&\displaystyle-\inf_{\pi\in \mathbb{R}^n}
 \bigg{\{} \partial_xV^{\mu}(t,x,y)[r(t,P_0(t))x+\gamma(t,P_0(t),P(t))\pi^{\top}]\\
& +\partial_yV^{\mu}(t,x,y)[r(t,P_0(t))y+\gamma(t,P_0(t),P(t))\pi^{\top}]\\
 &\displaystyle +\frac{1}{2} \partial^2_{xx}V^{\mu}(t,x,y)\pi\sigma(t,P(t))\sigma(t,P(t))^{\top}\pi^{\top} \bigg{\}},  \\
  \!\!\! V^{\mu}(T,x,y)=& \mu (x-y)^2-y,
\end{array}\right.
\end{equation}
where $\beta(t)=\gamma(t,P_0(t),P(t))[\sigma(t,P(t))\sigma(t,P(t))^{\top}]^{-1}\gamma(t,P_0(t),P(t))^{\top}$, and the related optimal strategy is
$$
\pi^*(t,x,y)=\frac{1}{2\mu}\gamma(t,P_0(t),P(t))[\sigma(t,P(t))\sigma(t,P(t))^{\top}]^{-1}
e^{-\int_t^Tr(h,P_0(h))\mathrm{d}h}.
$$
\end{theorem}

\begin{remark}\label{re-8}
Based on Remark \ref{re-0} and Theorem \ref{the-3}, we can obtain the time-consistent dynamic optimal strategy
$$
\pi^*(s)=\frac{1}{2\mu}\gamma(s,P_0(s),P(s))[\sigma(s,P(s))\sigma(s,P(s))^{\top}]^{-1}
e^{-\int_s^Tr(h,P_0(h))\mathrm{d}h}, \ t\leq s\leq T,
$$
which is independent from the state $(x,y)$ and the optimal value for cost functional is given as follows:
$$
V^{\mu}(t,x,x)=\lim_{y\to x}V^{\mu}(t,x,y)=-xe^{\int_t^Tr(h,P_0(h))\mathrm{d}h}
-\int_t^T\frac{\mathbb{E}[\beta(h)]}{4\mu}\mathrm{d}h,
$$
where the expectation $\mathbb{E}[\cdot]$ is based on the information of time $t$.

 In general, we can consider the following objective value function:
$$
V^{\mu}(t,x,y)=\inf_{\pi(\cdot)\in \mathcal{A}_t^T}\mathbb{E}[\Phi(X^{\pi}_{t,x}(T),\mathbb{E}[X^{\pi}_{t,y}(T)])],
$$
where $\Phi(x,y),\ x,y\in \mathbb{R}$ is a nonlinear function of $(x,y)$. We can obtain a Hamilton-Jocabi-Bellman equation for the value function $V^{\mu}(t,x,y)$ with boundary condition $V^{\mu}(T,x,y)=\Phi(x,y)$.
\end{remark}

\section{Conclusion}
To obtain a time-consistent dynamic optimal strategy for the classical continuous time mean-variance model, we view that the mean process $\mathbb{E}[X^{\pi}_{t,x}(\cdot)]$ should be recognized as a deterministic process that is different from the wealth process $X^{\pi}_{t,x}(\cdot)$. Then, we consider the following objective cost functional:
 \begin{equation}\label{c-cost}
\tilde{J}(t,x,y,\mu;\pi(\cdot))=\mu \mathbb{E}[\big(X^{\pi}_{t,x}(T)-Y^{\pi}_{t,y}(T)\big)^2]-Y^{\pi}_{t,y}(T).
\end{equation}
 From the cost functional (\ref{c-cost}), we can distinguish the wealth process $X^{\pi}_{t,x}(\cdot)$ and mean process $Y^{\pi}_{t,y}(\cdot)=\mathbb{E}[X^{\pi}_{t,y}(\cdot)]$ from the variance of the wealth. Based on these setting, we can derive a Hamilton-Jocabi-Bellman equation for the ternary value function $V^{\mu}(t,x,y)$. Our main results are given as follows:
\begin{itemize}
\item  A new method is proposed to deal with the objective cost functional when it contains a nonlinear part of the mean process $\mathbb{E}[X^{\pi}_{t,x}(\cdot)]$. This new method can help us to separate the nonlinear part of the mean process from the original objective cost functional.

 \item For the general setting, we can obtain the explicit formula for the value function $V^{\mu}(t,x,y)$. The time-consistent dynamic optimal strategy is found and is different from the existing results.

 \item Furthermore, the time-consistent relation of the mean and variance of this  mean-variance model is established.

\end{itemize}

\appendix
\section{The main proofs}\label{app}

\noindent {\textbf{Proof of Theorem \ref{the-1}.}}
Using the same technique in the proof of  Theorem 3.3, Chapter 4 in \cite{YZ99}, we can prove these results. For the reader's convenience, we show the main steps of this proof. In the following, for any given $0\leq t\leq s\leq T,\ x,y\in \mathbb{R}$, we set
$$
\tilde{V}^{\mu}(t,x,y)=\inf_{\pi(\cdot)\in \mathcal{A}_t^s}\mathbb{E}[V^{\mu}(s,X^{\pi}_{t,x}(s),Y^{\pi}_{t,y}(s))].
$$
By the definition of value function $V^{\mu}(t,x,y)$, for any given $\varepsilon>0$, there exists strategy $\tilde{\pi}(\cdot)$ (in the sense of weak formulation, see \cite{YZ99}) such that
\begin{equation}\label{app-hj-1}
\begin{array}{rl}
&{V}^{\mu}(t,x,y)+\varepsilon\\
\geq &\displaystyle  \mathbb{E}[\mu\big(X^{\tilde{\pi}}_{t,x}(T)-Y^{\tilde{\pi}}_{t,y}(T)\big)^2-Y^{\tilde{\pi}}_{t,y}(T)] \\
=&\displaystyle   \mathbb{E}\big[\mathbb{E}\big[\mu\big(X^{\tilde{\pi}}_{t,x}(T)-Y^{\tilde{\pi}}_{t,y}(T)\big)^2-Y^{\tilde{\pi}}_{t,y}(T)  \ |\ \mathcal{F}_s\big]\big]                           \\
=&\displaystyle   \mathbb{E}\big[\mathbb{E}
\big[\mu\big(X^{\tilde{\pi}}_{s,X^{\tilde{\pi}}_{t,x}(s)}(T)-Y^{\tilde{\pi}}_{s,Y^{\tilde{\pi}}_{t,y}(s)}(T)\big)^2-
Y^{\tilde{\pi}}_{s,Y^{\tilde{\pi}}_{t,y}(s)}(T) \ |\ \mathcal{F}_s\big]\big]                                           \\
=&\displaystyle  \mathbb{E}\big[
\tilde{J}(s,X^{\tilde{\pi}}_{t,x}(s),Y^{\tilde{\pi}}_{t,y}(s),\mu;\tilde{\pi}(\cdot)) \big]
\\
\geq& \displaystyle \mathbb{E}\big[V^{\mu}(s,X^{\tilde{\pi}}_{t,x}(s),Y^{\tilde{\pi}}_{t,y}(s))    \big]
\\
\geq& \tilde{V}^{\mu}(t,x,y).
\end{array}
\end{equation}
The third equality of (\ref{app-hj-1}) is derived by Lemma \ref{le-1}. In contrast, for the given $\varepsilon>0$, we want to prove ${V}^{\mu}(t,x,y)\leq  \tilde{V}^{\mu}(t,x,y)+\varepsilon$ in the following step. Based on Assumptions {$\textbf{H}_1$} and {$\textbf{H}_2$}, there exists $
\delta>0$ for any $\left|x_1-x_2\right|+\left|y_1-y_2\right|< \varepsilon$, we have that
$$
\left|\tilde{J}(t,x_1,y_1,\mu;\pi(\cdot))-\tilde{J}(t,x_2,y_2,\mu;\pi(\cdot))\right|+
\left|V^{\mu}(t,x_1,y_1)-V^{\mu}(t,x_2,y_2) \right|<\frac{\varepsilon}{3}.
$$
This inequality helps us find a strategy
\begin{eqnarray*}
\hat{\pi}(h)=\left\{\begin{array}{ll}
 \pi(h),\quad t\leq h\leq s,\\
\tilde{\pi}(h),\quad s<h\leq T,
\end{array}\right.
\end{eqnarray*}
where $\pi(\cdot)\in \mathcal{A}_t^s$ is a any given strategy, such that
$$
\tilde{J}(s,X^{{\pi}}_{t,x}(s),Y^{{\pi}}_{t,y}(s),\mu;\tilde{\pi}(\cdot))<
V^{\mu}(s,X^{{\pi}}_{t,x}(s),Y^{{\pi}}_{t,y}(s))+\varepsilon.
$$
Thus, for the strategy $\hat{\pi}(\cdot)$, we have
\begin{equation}\label{app-hj-2}
\begin{array}{rl}
&{V}^{\mu}(t,x,y)\\
\leq &\displaystyle
\mathbb{E}[\mu\big(X^{\hat{\pi}}_{t,x}(T)-Y^{\hat{\pi}}_{t,y}(T)\big)^2-Y^{\hat{\pi}}_{t,y}(T)] \\
=&\displaystyle   \mathbb{E}\big[\mathbb{E}\big[\mu\big(X^{\hat{\pi}}_{t,x}(T)-Y^{\hat{\pi}}_{t,y}(T)\big)^2-Y^{\hat{\pi}}_{t,y}(T)  \ |\ \mathcal{F}_s\big]\big]                           \\
=&\displaystyle   \mathbb{E}\big[\mathbb{E}
\big[\mu\big(X^{\hat{\pi}}_{s,X^{{\pi}}_{t,x}(s)}(T)-Y^{\hat{\pi}}_{s,Y^{{\pi}}_{t,y}(s)}(T)\big)^2-
Y^{\hat{\pi}}_{s,Y^{{\pi}}_{t,y}(s)}(T) \ |\ \mathcal{F}_s\big]\big]                                           \\
=&\displaystyle  \mathbb{E}\big[
\tilde{J}(s,X^{{\pi}}_{t,x}(s),Y^{{\pi}}_{t,y}(s),\mu;\hat{\pi}(\cdot)) \big]
\\
\leq & \displaystyle  \mathbb{E}\big[V^{\mu}(s,X^{{\pi}}_{t,x}(s),Y^{{\pi}}_{t,y}(s))\big]+\varepsilon,
\end{array}
\end{equation}
for $\pi(\cdot)\in \mathcal{A}_t^s$ is a any given strategy, we have
\begin{equation}\label{app-hj-3}
V^{\mu}(t,x,y)\leq \tilde{V}^{\mu}(t,x,y)+\varepsilon.
\end{equation}
Now, we combine equations (\ref{app-hj-1}) and (\ref{app-hj-3}) to obtain the equation (\ref{belm-1}). This completes the proof. $\quad \qquad \Box$

\bigskip

\noindent {\textbf{Proof of Theorem \ref{the-2}.}}
Note that, when $V^{\mu}(t,x,y)\in \mathcal{C}^{1,2,1}([0,T]\times\mathbb{R}\times\mathbb{R})$, we have that
\begin{equation*}
\begin{array}{rl}
0=&\displaystyle \inf_{\pi\in\mathcal{A}_t^s}\mathbb{E}[V^{\mu}(s,X^{\pi}_{t,x}(s),Y^{\pi}_{t,y}(s))
-V^{\mu}(t,x,y)]\\
=&\displaystyle \inf_{\pi\in\mathcal{A}_t^s}\mathbb{E}\bigg{[} V^{\mu}(t,x,y)(s-t)+\partial_xV^{\mu}(t,x,y)(X^{\pi}_{t,x}(s)-x)\\
&\displaystyle+\frac{1}{2}\partial^2_{xx}V^{\mu}(t,x,y)(X^{\pi}_{t,x}(s)-x)^2+\partial_yV^{\mu}(t,x,y)
(Y^{\pi}_{t,y}(s)-y)\bigg{]}+\mathrm{o}(s-t)\\
=&\displaystyle \inf_{\pi\in\mathcal{A}_t^s}\mathbb{E}\bigg{[} V^{\mu}(t,x,y)(s-t)+\partial_xV^{\mu}(t,x,y)(X^{\pi}_{t,x}(s)-x)\\
&\displaystyle+\frac{1}{2}\partial^2_{xx}V^{\mu}(t,x,y)(X^{\pi}_{t,x}(s)-x)^2+\partial_yV^{\mu}(t,x,y)
(X^{\pi}_{t,y}(s)-y)
\bigg{]}+\mathrm{o}(s-t),\\
\end{array}
\end{equation*}
the last equality is derived by the equation $Y^{\pi}_{t,y}(s)=\mathbb{E}[X^{\pi}_{t,y}(s)]$, where $\partial_tV^{\mu}(\cdot,\cdot,\cdot)$ means the partial derivative on time, while $\partial_xV^{\mu}(\cdot,\cdot,\cdot)$ and $\partial_yV^{\mu}(\cdot,\cdot,\cdot)$ mean the partial derivative on the first and second state of the value function $V^{\mu}(\cdot,\cdot,\cdot)$, respectively, and $\partial_{xx}^2V^{\mu}(\cdot,\cdot,\cdot)$ means the second-order partial derivative on the first state $x$. Dividing $s-t$ on both sides of this equation and letting $s\to t$, one obtains
 \begin{equation}\label{pde-1}
\left\{\begin{array}{rl}
  \!\!\! \partial_tV^{\mu}(t,x,y)=&\displaystyle-\inf_{\pi\in \mathbb{R}^n}
 \bigg{\{} \partial_xV^{\mu}(t,x,y)[r(t)x+\gamma(t)\pi^{\top}]
 +\partial_yV^{\mu}(t,x,y)[r(t)y+\gamma(t)\pi^{\top}]\\
 &\displaystyle+\frac{1}{2} \partial^2_{xx}V^{\mu}(t,x,y)\pi\sigma(t)\sigma(t)^{\top}\pi^{\top} \bigg{\}},  \\
  \!\!\! V^{\mu}(T,x,y)=& \mu (x-y)^2-y,\ 0\leq t\leq T.
\end{array}\right.
\end{equation}

In the first step, we assume $\partial^2_{xx}V^{\mu}(t,x,y)>0$; thus, the optimal strategy at time $t$ satisfies
$$
\pi^*(t,x,y)=\frac{\gamma(t)[\sigma(t)\sigma(t)^{\top}]^{-1}
[\partial_xV^{\mu}(t,x,y)+\partial_yV^{\mu}(t,x,y)]}
{-\partial^2_{xx}V^{\mu}(t,x,y)},
$$
which deduces that
 \begin{equation}\label{pde-2}
\begin{array}{rl}
  \!\!\! &\displaystyle
\partial_tV^{\mu}(t,x,y)+\partial_xV^{\mu}(t,x,y)r(t)x
 +\partial_yV^{\mu}(t,x,y)r(t)y\\
 =&-\displaystyle\frac{\beta(t)[\partial_xV^{\mu}(t,x,y)
 +\partial_yV^{\mu}(t,x,y)]^2}{2\partial^2_{xx}V^{\mu}(t,x,y)},
\end{array}
\end{equation}
where $\beta(t)=\gamma(t)[\sigma(t)\sigma(t)^{\top}]^{-1}\gamma(t)^{\top}$.

In the second step, we assume the solution to equation (\ref{pde-2}) is given as follows:
\begin{equation}\label{solu-1}
V^{\mu}(t,x,y)=A(t)(x-y)^2+B(t)y+C(t),
\end{equation}
where $A(\cdot),\ B(\cdot),\ C(\cdot)$ are the continuous derivable functions in $[0,T]$ with
$$
A(T)=\mu,\ B(T)=-1,\ C(T)=0.
$$
We plug the representation of $V^{\mu}(t,x,y)$ (\ref{solu-1}) into equation (\ref{pde-2}),
 \begin{equation}\label{pde-3}
\begin{array}{rl}
  \!\!\! &\displaystyle
A'(t)(x-y)^2+B'(t)y+C'(t)+2A(t)r(t)(x-y)x+2A(t)r(t)(y-x)y+B(t)r(t)y\\
 =&\displaystyle-\frac{\beta(t)[2A(t)(x-y)+2A(t)(y-x)+B(t)]^2}{4A(t)},
\end{array}
\end{equation}
then,
$$
[A'(t)+2A(t)r(t)](x-y)^2+[B'(t)+B(t)r(t)]y+C'(t)=-\frac{\beta(t)B(t)^2}{4A(t)}.
$$
Thus, we obtain the equations for $A(\cdot),\ B(\cdot), \ C(\cdot)$,
 \begin{equation}\label{pde-4}
\begin{array}{rl}
&\displaystyle
A'(t)+2A(t)r(t)=0,\ A(T)=\mu, \ 0\leq t\leq T;\\
& B'(t)+B(t)r(t)=0,\ B(T)=-1,\ 0\leq t\leq T;\\
&\displaystyle C'(t)=-\frac{\beta(t)B(t)^2}{4A(t)},\ C(T)=0,\qquad 0\leq t\leq T.\\
\end{array}
\end{equation}
 The solution to equation (\ref{pde-4}) is given as follows:
  \begin{equation}\label{pde-5}
\begin{array}{rl}
&\displaystyle
A(t)=\mu e^{\int_t^T2r(h)\mathrm{d}h}, \ 0\leq t\leq T;\\
& B(t)=-e^{\int_t^Tr(h)\mathrm{d}h},\ 0\leq t\leq T;\\
&\displaystyle C(t)=-\int_t^T\frac{\beta(h)}{4\mu}\mathrm{d}h,\ 0\leq t\leq T.\\
\end{array}
\end{equation}
Therefore, we have for $x\neq y$,
\begin{equation}\label{pde-6}
V^{\mu}(t,x,y)=\mu (x-y)^2e^{\int_t^T2r(h)\mathrm{d}h}-ye^{\int_t^Tr(h)\mathrm{d}h}
-\int_t^T\frac{\beta(h)}{4\mu}\mathrm{d}h.
\end{equation}
Notice that the risk aversion parameter $\mu>0$, thus, $\partial^2_{xx}V^{\mu}(t,x,y)>0$. The optimal strategy,
 \begin{equation}\label{pde-7}
\pi^*(t,x,y)=\frac{1}{2\mu}\gamma(t)[\sigma(t)\sigma(t)^{\top}]^{-1}
e^{-\int_t^Tr(h)\mathrm{d}h}, \ 0\leq t\leq T.
\end{equation}

Now, we can check the formula (\ref{pde-6}) of $V^{\mu}(t,x,y)\in \mathcal{C}^{1,2,1}([0,T]\times\mathbb{R}\times\mathbb{R})$ which is a classical solution to (\ref{pde-1}). Employing the uniqueness results from Theorem 6.1 Chapter 4 in \cite{YZ99}, we have that $V^{\mu}(t,x,y)$ in equation (\ref{pde-6}) is the unique classical solution of PDE  (\ref{pde-1}). This completes the proof. $\quad \qquad \Box$

\bigskip

\noindent {\textbf{Proof of Proposition \ref{pro-1}.}} For a given mean level $L>xe^{\int_t^{T}r(h)\mathrm{d}h}$ in constrained condition (\ref{mean-1}). The optimal strategy $\pi^*(\cdot)$ and $\pi^*_1(\cdot)$ satisfy
$$
\mathbb{E}[{X}^{{\pi}^{*}}_{t,x}(T)]=\mathbb{E}[{X}^{{\pi}^{*}_1}_{t,x}(T)]=L.
$$
By formulations (\ref{eff-1}) and (\ref{pre-eff}), we have
$$
\mathrm{Var}[X^{{\pi}^{*}}_{t,x}(T)]=
\frac{\bigg(L-xe^{\int_t^Tr(h)\mathrm{d}h}\bigg)^2}{\int_t^T\beta(h)\mathrm{d}h}, \quad \displaystyle  \mathrm{Var}[X^{{\pi}_1^{*}}_{t,x}(T)]=
\frac{\bigg{(}L
-xe^{\int_t^{T}r(h)\mathrm{d}h}\bigg{)}^2}{e^{\int_t^{T}\beta(h)\mathrm{d}h}-1}.
$$
By Assumption $\mathrm{\textbf{H}}_2$, we have $\beta(s)>0,\ t\leq s\leq T$, and
$$
\int_t^T\beta(h)\mathrm{d}h<e^{\int_t^{T}\beta(h)\mathrm{d}h}-1.
$$
Therefore, one obtains,
$$
\mathrm{Var}[X^{{\pi}^{*}}_{t,x}(T)]> \mathrm{Var}[X^{{\pi}_1^{*}}_{t,x}(T)].
$$

For a given risk aversion parameter $\mu>0$, we have
$$
\mathbb{E}[X^{{\pi}^{*}}_{t,x}(T)]=\displaystyle  xe^{\int_t^Tr(h)\mathrm{d}h}+\int_t^T\frac{\beta(h)}{2\mu}\mathrm{d}h,
$$
and
$$
\displaystyle \mathbb{E}[{X}^{{\pi}^{*}_1}_{t,x}(T)]=
x e^{\int_t^{T}r(h)\mathrm{d}h}+\frac{1}{2\mu}(e^{\int_t^{T}\beta(h)\mathrm{d}h}-1).
$$
From $\beta(s)>0,\ t\leq s\leq T$, it follows
$$
\int_t^T\frac{\beta(h)}{2\mu}\mathrm{d}h<\frac{1}{2\mu}(e^{\int_t^{T}\beta(h)\mathrm{d}h}-1),
$$
which implies that
$$
x e^{\int_t^{T}r(h)\mathrm{d}h}
<\mathbb{E}[X^{{\pi}^{*}}_{t,x}(T)]<\mathbb{E}[{X}^{{\pi}^{*}_1}_{t,x}(T)].
$$
Again, by formulations (\ref{eff-1}) and (\ref{pre-eff}), we have
$$
\mathrm{Var}[X^{{\pi}^{*}}_{t,x}(T)]=\frac{\int_t^T\beta(h)\mathrm{d}h}{4\mu^2}<
\frac{e^{\int_t^{T}\beta(h)\mathrm{d}h}-1}{4\mu^2}=\mathrm{Var}[X^{{\pi}_1^{*}}_{t,x}(T)].
$$
Therefore,
\begin{equation}\label{app-pro-1-var}
\mathrm{Var}[X^{{\pi}^{*}}_{t,x}(T)]<\mathrm{Var}[X^{{\pi}_1^{*}}_{t,x}(T)],\quad \mathbb{E}[{X}^{{\pi}^{*}}_{t,x}(T)]<\mathbb{E}[{X}^{{\pi}^{*}_1}_{t,x}(T)].
\end{equation}
This completes the proof. $\quad \qquad \Box$

\bigskip

\noindent {\textbf{Proof of Theorem \ref{the-3}.}
The proof of this Theorem is same with that in Theorem \ref{the-2}. For reader's convenience, we show the details of this proof. For any given $0\leq t\leq s\leq T,\ x,y\in \mathbb{R}$.
Using the technique in the proof of Theorem \ref{the-1}, we can obtain that
\begin{equation}\label{a-dpp}
V^{\mu}(t,x,y)=\inf_{\pi(\cdot)\in \mathcal{A}_t^s}\mathbb{E}[V^{\mu}(s,X^{\pi}_{t,x}(s),Y^{\pi}_{t,y}(s))].
\end{equation}
In the following, we assume $V^{\mu}(t,x,y)\in \mathcal{C}^{1,2,1}([0,T]\times\mathbb{R}\times\mathbb{R})$. Employing It\^{o} formula to $V^{\mu}(s,X^{\pi}_{t,x}(s),Y^{\pi}_{t,y}(s))$ and by equation (\ref{a-dpp}), it follows that
\begin{equation}\label{a-pde0}
\begin{array}{rl}
0=&\displaystyle \inf_{\pi\in\mathcal{A}_t^s}\mathbb{E}\bigg{[} V^{\mu}(t,x,y)(s-t)+\partial_xV^{\mu}(t,x,y)(X^{\pi}_{t,x}(s)-x)\\
&\displaystyle+\frac{1}{2}\partial^2_{xx}V^{\mu}(t,x,y)(X^{\pi}_{t,x}(s)-x)^2+\partial_yV^{\mu}(t,x,y)
(X^{\pi}_{t,y}(s)-y)
\bigg{]}+\mathrm{o}(s-t).\\
\end{array}
\end{equation}
Dividing $s-t$ on both sides of equation (\ref{a-pde0}) and letting $s\to t$, we have
 \begin{equation}\label{a-pde-1}
\left\{\begin{array}{rl}
  \!\!\! \partial_tV^{\mu}(t,x,y)=&\displaystyle-\inf_{\pi\in \mathbb{R}^n}
 \bigg{\{} \partial_xV^{\mu}(t,x,y)[r(t,P_0(t))x+\gamma(t,P_0(t),P(t))\pi^{\top}]\\
 &+\partial_yV^{\mu}(t,x,y)[r(t,P_0(t))y+\gamma(t,P_0(t),P(t))\pi^{\top}]\\
 &\displaystyle+\frac{1}{2} \partial^2_{xx}V^{\mu}(t,x,y)\pi\sigma(t,P(t))\sigma(t,P(t))^{\top}\pi^{\top} \bigg{\}},  \\
  \!\!\! V^{\mu}(T,x,y)=& \mu (x-y)^2-y,\ 0\leq t\leq T.
\end{array}\right.
\end{equation}

In addition, we assume $\partial^2_{xx}V^{\mu}(t,x,y)>0$; thus, the optimal strategy at time $t$ satisfies
$$
\pi^*(t,x,y)=\frac{\gamma(t,P_0(t),P(t))[\sigma(t,P(t))\sigma(t,P(t))^{\top}]^{-1}
[\partial_xV^{\mu}(t,x,y)+\partial_yV^{\mu}(t,x,y)]}
{-\partial^2_{xx}V^{\mu}(t,x,y)},
$$
and
 \begin{equation}\label{a-pde-2}
\begin{array}{rl}
  \!\!\! &\displaystyle
\partial_tV^{\mu}(t,x,y)+\partial_xV^{\mu}(t,x,y)r(t,P_0(t))x
 +\partial_yV^{\mu}(t,x,y)r(t,P_0(t))y\\
 =&-\displaystyle\frac{\beta(t)[\partial_xV^{\mu}(t,x,y)
 +\partial_yV^{\mu}(t,x,y)]^2}{2\partial^2_{xx}V^{\mu}(t,x,y)},
\end{array}
\end{equation}
where $\beta(t)=\gamma(t,P_0(t),P(t))[\sigma(t,P(t))\sigma(t,P(t))^{\top}]^{-1}\gamma(t,P_0(t),P(t))^{\top}$.

In the following, we assume the solution to equation (\ref{a-pde-2}) is given as follows:
\begin{equation}\label{a-solu-1}
V^{\mu}(t,x,y)=A(t)(x-y)^2+B(t)y+C(t),
\end{equation}
where $A(\cdot),\ B(\cdot),\ C(\cdot)$ are the continuous derivable functions in $[0,T]$ with
$$
A(T)=\mu,\ B(T)=-1,\ C(T)=0.
$$
We plug the representation of $V^{\mu}(t,x,y)$ (\ref{a-solu-1}) into equation (\ref{a-pde-2}). Then, we can obtain the equations for $A(\cdot),\ B(\cdot), \ C(\cdot)$,
 \begin{equation}\label{a-pde-4}
\begin{array}{rl}
&\displaystyle
A'(t)+2A(t)r(t,P_0(t))=0,\ A(T)=\mu, \ 0\leq t\leq T;\\
& B'(t)+B(t)r(t,P_0(t))=0,\ B(T)=-1,\ 0\leq t\leq T;\\
&\displaystyle C'(t)=-\frac{\beta(t)B(t)^2}{4A(t)},\ C(T)=0,\qquad 0\leq t\leq T.\\
\end{array}
\end{equation}
Notice that, for $s>t$, $\beta(s)$ is a random variable. To find an adapted solution for $C(\cdot)$, we take the expectation $\mathbb{E}[\cdot]$ on both sides of the third equation of (\ref{a-pde-4}), where the expectation $\mathbb{E}[\cdot]$ is based on the information of time $t$. The solution to equation (\ref{a-pde-4}) is given as follows:
  \begin{equation}\label{a-pde-5}
\begin{array}{rl}
&\displaystyle
A(t)=\mu e^{\int_t^T2r(h,P_0(h))\mathrm{d}h}, \ 0\leq t\leq T;\\
& B(t)=-e^{\int_t^Tr(h,P_0(h))\mathrm{d}h},\ 0\leq t\leq T;\\
&\displaystyle C(t)=-\int_t^T\frac{\mathbb{E}[\beta(h)]}{4\mu}\mathrm{d}h,\ 0\leq t\leq T.\\
\end{array}
\end{equation}
Therefore, we have
\begin{equation}\label{a-pde-6}
V^{\mu}(t,x,y)=\mu (x-y)^2e^{\int_t^T2r(h,P_0(h))\mathrm{d}h}-ye^{\int_t^Tr(h,P_0(h))\mathrm{d}h}
-\int_t^T\frac{\mathbb{E}[\beta(h)]}{4\mu}\mathrm{d}h.
\end{equation}
 Notice that the risk aversion parameter $\mu>0$, thus, $\partial^2_{xx}V^{\mu}(t,x,y)>0$. The optimal strategy is given as follows:
 \begin{equation}\label{a-pde-7}
\pi^*(t,x,y)=\frac{1}{2\mu}\gamma(t,P_0(t),P(t))[\sigma(t,P(t))\sigma(t,P(t))^{\top}]^{-1}
e^{-\int_t^Tr(h,P_0(h))\mathrm{d}h}, \ 0\leq t\leq T.
\end{equation}
The following proof is same with that in Theorem \ref{the-2}. Thus, we omit it. This completes the proof. $\quad \qquad \Box$

\newpage
\bibliography{var}
\end{document}